\documentclass[nonacm,sigconf]{acmart}
\AtBeginDocument{%
  \providecommand\BibTeX{{%
    \normalfont B\kern-0.5em{\scshape i\kern-0.25em b}\kern-0.8em\TeX}}}

\setcopyright{rightsretained}
\copyrightyear{2023} 
\acmYear{2023}
\setcopyright{rightsretained}
\acmConference[UIST '23 Adjunct]{The Adjunct Publication of the 36th Annual ACM Symposium on User Interface Software and Technology}{Oct.29-Nov.1, 2023}{San Francisco,CA, USA} 
\begin{document}
%\title[Memory Sandbox]{Memory Sandbox: Transparent and Manipulable Memory Management for Conversational Agents}
\title[Memory Sandbox]{Memory Sandbox: Transparent and Interactive Memory Management for Conversational Agents}

\author{Ziheng Huang}
\email{z8huang@ucsd.edu}
\affiliation{%
  \institution{University of California---San Diego}
  \streetaddress{9500 Gilman Drive}
  \city{San Diego}
  \state{CA}
  \country{USA}
  \postcode{92093}
}

\author{Sebastian Gutierrez}
\email{guts@temple.edu}
\affiliation{
  \institution{Temple University}
  \streetaddress{1801 N Broad St}
  \city{Philadelphia}
  \state{PA}
  \country{USA}
  \postcode{19122}
}

\author{Hemanth Kamana}
\email{tuo73589@temple.edu}
\affiliation{
  \institution{Temple University}
  \streetaddress{1801 N Broad St}
  \city{Philadelphia}
  \state{PA}
  \country{USA}
  \postcode{19122}
}

\author{Stephen MacNeil}
\email{stephen.macneil@temple.edu}
\affiliation{
  \institution{Temple University}
  \streetaddress{1801 N Broad St}
  \city{Philadelphia}
  \state{PA}
  \country{USA}
  \postcode{19122}
}

\renewcommand{\shortauthors}{Huang, et al.}

\begin{abstract}

The recent advent of large language models (LLM) has resulted in high-performing conversational agents such as chatGPT. These agents must remember key information from an ongoing conversation to provide responses that are contextually relevant to the user. However, these agents have limited memory and can be distracted by irrelevant parts of the conversation. While many strategies exist to manage conversational memory, users currently lack affordances for viewing and controlling what the agent remembers, resulting in a poor mental model and conversational breakdowns. In this paper, we present Memory Sandbox, an interactive system and design probe that allows users to manage the conversational memory of LLM-powered agents. By treating memories as data objects that can be viewed, manipulated, recorded, summarized, and shared across conversations, Memory Sandbox provides interaction affordances for users to manage how the agent should `see' the conversation.

%Intelligent conversational agents rely on remembering key information from an ongoing conversation to provide responses that are contextually appropriate. However, agents that are powered by large language models have limited memory and can be distracted by irrelevant conversational context. While many strategies exist to manage conversational memory, end users lack transparency and control over what and how memory is managed by the agent, resulting in a poor mental model of what parts of the conversation the agent uses or remembers. In this paper, we present Memory Sandbox, an interactive system and design probe that allows end users to see and manage the conversational memory of agents. By treating memories as objects that can be manipulated, reordered, summarized, and shared across conversations, Memory Sandbox provides interaction affordances for end users to manage how the agent should 'see' the conversation.

% conversational memory of agents to align with user understanding of the conversation. 

% Memory Sandbox also allows users to reorder and share memories across multiple conversations. 

% to align with their intent. Memory Sandbox provides interaction affordances for users to manipulate the visibility and representation of memory. Users can also interact with the memory by dragging and dropping 

\end{abstract}


%\ccsdesc[500]{Human-centered computing~Visualization}
% \ccsdesc[500]{Applied computing~Document management and text processing}
% \ccsdesc[500]{Human-centered computing~Collaborative and social computing}
%\ccsdesc[500]{Information systems~Information systems applications}


%, Conversational Agents} 

\maketitle

\section{Introduction}

Large Language Models (LLMs) are currently capable of generating human-like responses in open-domain tasks~\cite{brown2020language}. This has led to a new generation of conversational agents, such as chatGPT, which are now being widely used across domains. 
% As a result, conversational interfaces/agents powered by LLMs such as ChatGPT started to emerge.
To ensure that agents generate responses that are contextually relevant and coherent to an ongoing conversation, these agents must maintain a working memory of the conversational history that has occurred up to that point in the conversation. 
The default strategy is to use as much of the conversational history as will fit within the input size limit of the LLM. Parts of the conversations that go beyond that buffer limit are forgotten, which leads to breakdowns when users assume the model remembers past context. 
% For example, GPT-4 has a context window limit of 32K tokens. 
% The default strategy for representing and using conversational memory is to leverage a \textit{memory buffer} which stores the last K tokens. For example, GPT-4 has a context window/buffer limit of 32K tokens. 
Additionally, as the input buffer size increases, the performance of the LLM degrades as it struggles to retrieve relevant context and can be distracted by irrelevant context~\cite{liu2023lost,shi2023large}.  
% regardless of the models' context window limit~\cite{liu2023lost}. LLMs can also be distracted by irrelevant context~\cite{shi2023large}. 
% As the conversation continues, LLM-powered agents can produce unexpected output~\cite{herdingaicat}. 
This problem is compounded because users do not know how the LLM is leveraging the memory to generate responses.%large stack of memory to generate responses. 
% }

Multiple strategies have been introduced to manage agents' conversational memory. For example, the conversation can be automatically summarized~\cite{xu2021beyond} and refined~\cite{zhong2022less} to reduce redundancy while maintaining key information. Some systems selectively store~\cite{xu-etal-2022-long,ma2021one} and update~\cite{bae2022keep} key memories. Relevant memories can also be retrieved based on the user input~\cite{xu2021beyond,bae2022keep,park2023generative}. 
% and memory retrieval approaches automaticly retrieve   relevant memories to present to the agent~\cite{}. 
% Memory retrieval approaches allows agents to retrieve relevant memories to the given prompt from the entire past histories ~\cite{xu2021beyond,bae2022keep}. 
% Memories can also be updated as the conversation proceeds~\cite{bae2022keep}.
% More complex memory models leveraging a combination of these strategies also exist~\cite{}. 
However, these memory management strategies are hidden behind the interface, resulting in a lack of transparency. Users often do not know what strategy is being used and have limited control over it. This makes it difficult for users to repair conversational breakdowns that happen when there is a misalignment between how the agent manages the memory and how the user perceives the conversation. 

\begin{figure*}
  \includegraphics[width=1\textwidth]{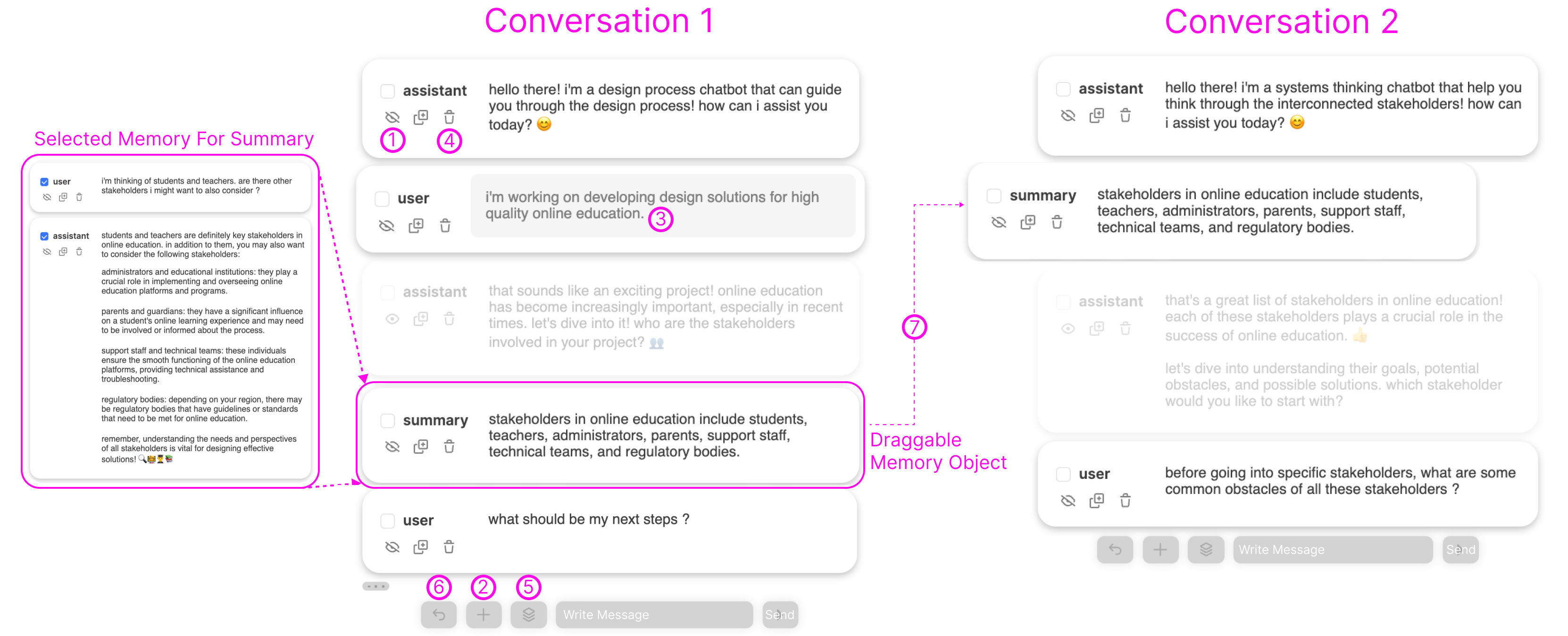}
\caption{Memory Sandbox is a system that enables users to see and manage the memory of conversational agents. Memory Sandbox provides the following interaction affordances: 1) toggle memory visibility, 2) add memory, 3) edit memory, 4) delete memory, 5) summarize memory, 6) create a new conversation, and 7) share memory.}
 % \Description{A website is presented with three main areas. The first is an area for writing problem statements. The second is an area for selecting stakeholders. The third is an area for viewing problem statements.}
  \label{fig:teaser}
\end{figure*}

% - memory models are not explainable, there are limits
% - ppl don't know the limits, don't have a mental model
% - leads to conversational breakdowns 
% - some people are able to partially mitigate with folk theories, experimentation, and exploration/sensmaking.

We present \textit{Memory sandbox}, shown in Figure~\ref{fig:teaser}, a system that allows users to see and manage the memory of conversational agents to align with user understanding of the conversation. Memory Sandbox transforms conversational memory, previously managed behind the user interface, into \textbf{interactive memory objects} within the interface. Users can manipulate the visibility and content of memory objects, spatially rearrange them, and share them across conversations.  %Memory Sandbox allows users to manipulate the visibility and representation of memory objects. Users can also drag memory objects for spacial rearrangement or memory sharing across conversations. 
% We make the following contributions: 1) a technique to make conversational memory management transparent and manipulable with interactive memory objects and 2) The Memory Sandbox system that offers novel interaction affordances for users to see and manipulate conversational memory to manage how the agent should 'see' the conversation. 
We make the following contributions: 1) The conceptualization of memory objects which makes conversational memory transparent and interactive and 2) The Memory Sandbox system that offers novel interaction affordances for users to view and manipulate the conversational memory of an intelligent agent.

% mutable. 
% manipulable

% the concept of transforming previously intractable conversational memory into interactive memory objects, 2) 

\section{System Overview}

\textit{Memory sandbox} is a system that provides users with the ability to view and manipulate the memory model of an intelligent agent, resulting in a shared representation of their ongoing conversation. Memory Sandbox introduces the concept of a \textit{memory object}, an interactive piece of conversational history that can be moved, edited, deleted, or combined with other memory objects through summarization. The interface is implemented in Next.js and uses the GPT-3.5 turbo model from the OpenAI API. Below we present the features of Memory Sandbox to help end users view and manage an LLM-powered agent's memory model.% represented with memory objects.
% manipulate conversational memory visibility and curate memory representation.

% \begin{itemize}    
%     \item \textbf{Hide/Show Memory Unit} 
%     \item \textbf{Reorder Memory Unit} 
%     \item \textbf{Edit Memory Unit} 
%     \item \textbf{Delete Memory Unit} 
%     \item \textbf{Add Memory Unit} 
%     \item \textbf{Summarize Memory Units} 
% \end{itemize}
% view and manipulate 
%\subsection{Transparent and interactive memory object}
\subsection{View and manipulate memory objects}
% prev system provide no control. for those that allow control, existing controls not enough. 
% Conversational memory of agents can't currently be manipulated by the user to explicitly align with their intent. 

% Yet, memory is being treated as an ordered list which limits the flexibility and expressiveness in allowing users to manipulate the memory to align with their intent.
% currently available interaction is limited in flexibility and expressiveness in allowing users to manipulate the memory to align with their intent.
% memory is treated as an ordered list which limits the flexibility and expressiveness of the interaction and the ability for users to manipulate the memory to align with their dynamic intent.
% The conversational memory of agents is managed behind the scenes, resulting in a misalignment between the full conversational history displayed on the frontend with the managed memory on the backend. Memory Sandbox fills this gap by exposing the managed memory on the front end. What users see is aligned with what the agent sees. 
{\color{red}
% The conversational memory of agents was previously managed behind the interface.
% End users currently don't have access and control over how agents manage conversational memory. Memory Sandbox probes the design of interaction affordances for end users to see and manipulate agents' memory.
}
% Prior research has explored how objectifying tools~\cite{bier1993toolglass, bederson1996local} and attributes~\cite{xia2016object} enable flexible, expressive, and cognitively direct manipulation of the interface. 
% Additionally, how static `bits' can become tangible and interactive~\cite{ishii1997tangible}. 

% {\color{blue}
Explainable AI research seeks to help people form mental models of intelligent systems~\cite{rutjes2019considerations}. Transparency of the inner workings of the system~\cite{zhang2020effect,eiband2018bringing} and interactivity to probe and manipulate the system~\cite{ross2021evaluating} have been demonstrated to help people interpret and interact with intelligent systems to achieve their goals.
% }
% , the concepts of transparency~\cite{} and interactivity~\cite{interactivereconstruction} were introduced. 
% Transparency provides insight into how a response is generated from the system. On the other hand, interactivity allows users to 

%Additionally, prior research has explored how objectifying tools~\cite{bier1993toolglass, bederson1996local} and attributes~\cite{xia2016object} enable flexible, expressive, and cognitively direct manipulation of the interface. 

% make opaque algorithmtic/AI systems transparent and understandable to end users. 

% Extensive works exist in explainibility 

% enabling people’s understanding of the AI to achieve their goals.

%To facilitate transparency and the formulation of mental models, 
Memory Sandbox makes the conversational memory explicit through the use of `memory objects' which can be viewed and manipulated within the interface. This was inspired by prior work that `objectifies' tools~\cite{bier1993toolglass, bederson1996local} and attributes~\cite{xia2016object} to enable flexibility, expressiveness, and direct manipulation. % of the interface. 
This results in a `shared representation'~\cite{horvitz1999principles, heer2019agency} and common ground~\cite{clark1989contributing}---so what users see on the front-end is what an LLM would `see' on the back-end. 

%Memory Sandbox exposes and objectifies the conversational memory that was previously managed implicitly behind the interface into memory objects that can be directly manipulated within the interface. This takes inspiration from prior work on `objectifying' tools~\cite{bier1993toolglass, bederson1996local} and attributes~\cite{xia2016object} to enable flexible, expressive, and direct manipulation of the interface. This also provides a `shared representation'~\cite{horvitz1999principles, heer2019agency} and common ground~\cite{clark1989contributing}---so what users see on the front-end is what an LLM would `see' on the back-end. 
%What users see on the front end is aligned with what the agent sees on the back end. 
%Each Memory object represents a single message of past conversation from the agent, the user, or a systems message. 
Additionally, users can view, edit, add, and delete memory objects to directly control how the agent 'sees' the conversation.  
% This idea of shared rep is a core conponant for success mix-init sys, facilitate model transparency, an important topic for explainable ai. 

% These independently manipulable memory objects provide fine-grained control for users to manage memory and how the agent should 'see' the conversation. 
% by manipulating the visibility and representation of conversational memory. 

% transparency, saliency map

\subsection{Toggle memory object visibility}
% What context is being presented to the model ?
% {\color{red} 
% LLMs infer meaning from conversational context and as conversations unfold there is more potential for LLMs to be distracted by irrelevant details  
% %and more of the conversational context needs `remembered' by the model
% ~\cite{liu2023lost,shi2023large}. 
% However, users currently do not have adequate affordances for viewing what is remembered or forgotten by LLMs from a large amount of input context. 

% Users currently don't have access to how memory is managed by LLM-powered agents. 

As a conversation grows, LLMs must increasingly rely on their memory management strategy to infer meaning from the conversation. However, in longer conversations, it is unclear what parts of the conversation are stored in memory or are attended to by the model~\cite{liu2023lost}. This results in a poor mental model for users and a lack of control over what context is maintained and used by the agent.
% } 

% Memory management strategies that more selectively present memory are hidden behind the scenes~\cite {}. Users don't have a clear mental model and explicit control over what context is being shared with the model. 
% This problem can be magnified when research have demonstrated that people engage in exploratory behavior with context switching when interacting with LLMs~\cite{}. 

% intent can alter and evolves throughout the conversation. 

% They explore new topics but also go back to past topics. To support this back-and-force interaction, MS 

Memory Sandbox enables users to selectively hide or show memory objects to control what context is shared with the agent. When the user's intent changes or the conversational context switches, the user can toggle the visibility of memory objects to hide or show parts of the conversation. %to update what context is being shared with the agent. 
%To signify the memory objects that are visible to the model, 
As a signifier, hidden memory objects are grayed out within the interface. 
% Additionally, exploratory hiding and showing memory objects provide a way for users to unit test how specific memory objects influence LLM output.
% and how different framings of the same memory object effects LLM output. 

% reflection curation, sensemaking, concept drift.

% theories of conversation/communication
% iterative conversational grounding.

% \subsection{Manipulating memory representation (add, edit, delete, summarize)}
\subsection{Curate memory objects}
% How the context is being presented to the model?
% LLMs are extremely sensitive to input contexts. Yet, important contexts in a conversation are often incomplete and scattered. Thus, in addition to allowing users to manipulate the visibility of memory objects, Memory Sandbox allows users to curate memory objects that are presented to the model. 
Discussants develop and refine their understanding as a conversation unfolds~\cite{clark1989contributing}. %Thus, in addition to allowing users to manipulate the visibility of memory objects, 
Thus, Memory Sandbox provides controls for users to curate memory objects %as the conversation evolves. 
%The user can 
by editing an existing memory object to refine or update the context, deleting a memory object to remove completely irrelevant context, and adding a new memory object to supplement extra context.
% or experiment with a different prompt/framing of existing memory objects
% these affordance exist for people, ...
Additionally, the arrangement of context is shown to have a significant effect on how well LLMs are able to leverage relevant context~\cite{liu2023lost}. In Memory Sandbox, all the memory objects are draggable, allowing users to experiment and refine the ordering and placement of memory objects in a conversation. 

\subsection{Summarize memory objects}

% Important contexts are often incomplete and scattered throughout the conversation~\cite{}. 

Reminiscent of how humans attend to key aspects in a conversation~\cite{mccarthy1991experimental}, abstractive summarization distills a large amount of information to provide essential elements to the agent. Yet, what is considered as `key aspects' can vary for individuals, even in the same conversation~\cite{mccarthy1991experimental}. Memory Sandbox enables uses to select memory objects that are summarized by the LLM. The resulting memory object represents the previous conversation and can be further refined by the user. The original conversation can be viewed by clicking on the summary.
%The newly created memory object containing the summary serves as a starting point to be further refined. Users have the control to edit the summary and manipulate the prompt used to obtain the summary.

\subsection{Share memory objects across conversations}

% LLM-powered conversational agents have been criticized for their inherently linear conversational interaction. While existing interfaces such as chatgpt allow users to create new conversations to strategically assign subtasks, 

Aligning with the goal of managing memory, Memory Sandbox also provides affordances for sharing memories across conversations. This offers a new way for users to engage with multiple agents outside of a single conversation thread. 
% This provides interesting opportunities for mult-agent interaction because 
Unlike in conversations with people, the speaker doesn't need to repeat themselves in each conversation to establish a shared understanding. 
% @fred, don't make this sound like two different things! See above
%In addition to providing interaction affordances to help end users curate conversational memory within a single conversation, Memory Sandbox probes the design space of multi-agent interaction. Often, we discuss ideas with multiple people but we need to repeat ourselves in each conversation to establish common understanding.

% Existing interfaces such as chatgpt allow users to create new conversations where the user to move to other conversations by choosing from the menu. Yet, going back and force across conversations can be tedious ......

Users can create and start multiple conversations with separate LLM-powered agents in the same 2D canvas. Memory objects can be shared and connected between conversations by dragging the memory object from one conversation to another. When dragging, memories are copied by reference to help the user identify the context source. 

%Memory Sandbox seeks to optimize the interaction when multiple conversational agents are involved. Users can create multiple conversational agents side by side on the same screen. Memory objects are not confined to a single conversation. Users can share memory objects across agents by dragging the source memory object to the target conversation, allowing users to control what memory objects are shared across agents. To keep the original conversation unchanged, users can also use the duplicate function to create a copy of the source memory object which can be dragged to the target conversation.

% \subsection{Reordering and sharing memory / Interacting with memory (drag/drop)}

% motivations:....

% By transforming intractable conversational memory into interactive memory objects, memory objects can be dragged within or across conversations. Within the same conversation, the user can drag the to reorder the memory objects for structural and process revision. Additionally, Memory Sandbox allows users to create multiple conversations side by side. Users can share memory objects across conversations by dragging or dropping. 

% Similar to interfaces such as gpt, memory sandbox allows users to create multiple conversations. As an interactive unit, Memory objects can also be shared across conversations with dragging and dropping.  

% add, edit, summary, delete, reorder       

% \section{System case study}
% To demonstrate Memory Sandbox's capabilities, we present a case study of Fred who is developing a proposal for 'making online education more interactive' with a design chatbot. 

\section{Discussion} %discussion and future work

% Fred I sketched this discussion based on what you had... 

% Conversations break down when the model doesn't have enough (or the right) context 
% The user has a poor mental model of what context the model "sees" and uses 
% Our tool supports transparency and enables users to form a mental model 
% With this aided transparency and explainability, they are able to curate their conversation and engage in conversational repair as needed. 
% Future work can explore how people engage with this tool. 

% Conversational repair
% Conversational curation 
% Transparency and explainability 
% Limitations: even if memory sizes continue to increase, this kind of tool gives users agency over the memory model which may reduce the need for models to infer. By hiding irrelevant details, “pinning” important aspects, etc. We can give users more control over how the bot “sees” the conversation.

Conversing is a collaborative activity where participants develop common ground through summarizing the discussion, repairing breakdowns, and emphasizing or de-emphasizing shared ideas~\cite{clark1989contributing}. Yet, existing chatbot interfaces do not provide affordances for understanding how the agent `sees' the conversation. Additionally, users can not rely on a theory of mind. These aspects result in a poor mental model for users and potential misalignment in understanding where conversational breakdown can occur.%end users currently have a poor mental model and control of what context the conversational agent “sees” and how the context is leveraged. When a misalignment between agent understanding and user understanding occurs, conversation breakdown can occur. 

Memory Sandbox transforms previously implicitly managed conversational memory behind the interface into interactive memory objects on the interface, exposing full control over the memory model of the agent to end users. 
% By hiding irrelevant details, `pinning' important aspects,
By selectively hiding, showing, and curating memory representation, we can give users more control over how the agent should “see” the conversation. In addition to curating memory in a single conversation, Memory Sandbox is also a design probe toward memory manipulation affordances for multi-agent interactions. By displaying multiple agents on the same screen and making memories interactive and draggable, Memory Sandbox allows end users to selectively control the shared or unique memory each agent contains. 

% {\color{blue}
Tools are beginning to emerge that focus on how users might interact with LLMs, including mapping UI affordances to an LLM~\cite{macneil2023prompt}, grounding human-AI collaboration in a shared artifact~\cite{huang2023causalmapper}, providing templates to facilitate prompt generation~\cite{jiang2022promptmaker}, and decomposing complex prompts to facilitate debugging~\cite{wu2022ai}. In this paper, we presented Memory Sandbox an interactive system that probes the design space of interaction techniques for memory management of LLMs. Our future work includes user studies to evaluate the efficacy of these techniques and potential trade-offs for implicit vs explicit memory management
% }

% Previous works have explored extensively supporting the prototyping of a single prompt and its difficulties~\cite{}. Memory Sandbox provides affordances for users to manipulate and prototype conversational agents' memory.
% introduces the concept of memory prototyping where users are given the control to prototype and curate one or more agent’s memory throughout the conversation. 
% A series of user studies will be performed to evaluate the difficulties and benefits of memory prototyping and how affordances offered by Memory Sandbox support memory prototyping.

\balance
\bibliographystyle{ACM-Reference-Format}
\bibliography{MemorySandbox}


\end{document}